\documentclass[aps,pra,twocolumn,showpacs,superscriptaddress]{revtex4}
\usepackage{amssymb,graphics,graphicx,times,bm,bbm,color}

\begin{document}

\title{Role of quantum coherence in chromophoric energy transport}
\author{Patrick Rebentrost}
\affiliation{Department of Chemistry and Chemical Biology, Harvard University, 12 Oxford
St., Cambridge, MA 02138}
\author{Masoud Mohseni}
\affiliation{Department of Chemistry and Chemical Biology, Harvard University, 12 Oxford
St., Cambridge, MA 02138}
\author{Al\'an Aspuru-Guzik}
\affiliation{Department of Chemistry and Chemical Biology, Harvard University, 12 Oxford
St., Cambridge, MA 02138}

\begin{abstract}
The role of quantum coherence and the environment in the
dynamics of excitation energy transfer is not fully understood. In
this work, we introduce the concept of dynamical contributions of
various physical processes to the energy transfer efficiency. We
develop two complementary approaches, based on a Green's function
method and energy transfer susceptibilities, and quantify the
importance of the Hamiltonian evolution, phonon-induced decoherence, and
spatial relaxation pathways. We investigate the Fenna-Matthews-Olson
protein complex, where we find a contribution of coherent dynamics
of about 10\% and of relaxation of 80\%.

\end{abstract}

\volumeyear{year}
\volumenumber{number}
\issuenumber{number}
\eid{identifier}
\date{ \today }
\startpage{1}
\maketitle

\section{Introduction}

Exciton transfer among chlorophyll molecules is the energy transport mechanism
of the initial step of the photosynthetic process. Light is captured by an
antenna complex and the exciton is subsequently transferred to a reaction center
where bio-chemical energy storage is initiated by a charge separation event
\cite{Blankenship02}. This transfer process has been studied using classical
F\"{o}rster theory or a (modified) Redfield/Lindblad description
\cite{Forster65,GroverSilbey71,Damjanovi97,YangFleming02,Scholes03,Grondelle04,MayBook,Jang04,Mohseni08}.
Several measures such as the energy transfer efficiency/quantum yield, transfer time,
and exciton lifetime have been employed to elucidate the performance of
exciton transfer \cite{Sener04,Leegwater96,Castro07}. Recent experiments suggest evidence of
long-lived quantum coherence in the Fenna-Matthews-Olson (FMO) protein complex
of the Green-Sulphur bacterium \textit{Chlorobium tepidum} and in the reaction
center of the purple bacterium \textit{Rhodobacter sphaeroides} \cite%
{Engel07, Lee07}. A naturally arising question is the role of coherence in the
biological function of the aforementioned chromophoric complexes.

In this work, we investigate relevant quantum coherence effects by an
\textit{in-situ} analysis of a success criterion for the initial step
in photosynthesis, the energy transfer efficiency (ETE).
The dynamics of an excitation in multi-chromophoric complexes can be
described in terms of an environment-assisted quantum walk \cite{Mohseni08}:
\begin{equation}
\frac{d}{d t}\rho(t)=\mathcal{M}\rho(t).
\label{LiouvilleQuantumWalk}
\end{equation}
The evolution generated by the superoperator $\mathcal{M}$ connects
the population and coherence elements of the density matrix $\rho$.
We assume in this paper that $\mathcal{M}$ is time-independent.
We would like to explore and characterize the dynamics of
Eq.~(\ref{LiouvilleQuantumWalk}), especially the role of quantum coherence,
the environment, and spatial energy transfer pathways.
For closely packed multichromophoric arrays, such as the
FMO complex, one has to account for strong inter-molecular
coupling and quantum coherence effects.
A classical approximation of the master
equation would be an insufficient description of the open quantum dynamics.
Specifically, we want to avoid any comparison
of the actual open quantum system under study
with a fictitious or abstract model system, such as a
high temperature limit, a strong decoherence model, or a semi-classical
F\"{o}rster method.
This is in contrast to studies e.g.~in the area of quantum information
that compare the quantum dynamics to classical dynamics, for example in the case of
the comparison of a classical random walk to a quantum walk \cite{Childs02}.

We quantify the role of the various physical processes
involved in the energy transfer process in terms of their contribution
to the ETE.
Formally, we will partition the overall
ETE, $\eta$, into a sum of terms,
\begin{equation} \label{EquationPartitioning}
\eta =\sum_{k}\eta _{k},  \label{EfficiencyContributionSum}
\end{equation}%
corresponding to a physical decomposition
$\mathcal{M=}\sum_{k}\mathcal{M}_{k}$. Each term $\eta _{k}$ can be interpreted as a
contribution to the overall efficiency originating from a
particular process $\mathcal{M}_{k}$.
For example, we will split the superoperator into the major components describing the exciton dynamics:
coherent evolution with the excitonic Hamiltonian, relaxation within the single-exciton manifold, and dephasing.
The $\eta_k$ associated with the coherent part will then give an indication
of the role of quantum evolution to the energy transfer efficiency and hence
to the biological function within a particular chromophoric complex.
We note that the exact partitioning of the ETE
into a sum of terms like Eq.~(\ref{EquationPartitioning}) is a non-trivial task:
as will be described below, the ETE essentially involves an exponential mapping
of the complete superoperator $\mathcal{M}$.
A separation of the ETE into a \textit{product} of terms
would seem more natural but would not allow the interpretation of $\eta_k$ as contributions.

In the following sections, we briefly
discuss the structure of the superoperator $\mathcal{M}$ and introduce two
complementary measures of efficiency contributions: one is  based on a Green's
function method and the other is derived from  energy transfer susceptibilities.
We apply these two approaches to the study of the ETE in the FMO complex.
We employ a standard Redfield model with the secular approximation which leads to
a master equation in Lindblad form \cite{BreuerBook}. This model captures major
decoherence effects such as relaxation and dephasing. We also include spatial correlation
of the fluctuations. The Markovian approximation neglects temporal
correlations in the phonon bath which can be relevant in photosynthetic systems
and will be treated in subsequent work.
We believe that this model and our methods can provide insight into the role
of quantum coherence in energy transfer, a process that occurs in noisy,
ambient temperature environments.

\section{Master equation for multichromophoric systems}

The transport dynamics of a single excitation is described by a master equation
for the density matrix that includes coherent evolution, relaxation, and
dephasing. Moreover, the exciton can recombine or be trapped.
The Hamiltonian for an interacting $N$-chromophoric system in the presence of a single
excitation can be written as \cite{MayBook}:
\begin{equation}
H_{\mathrm{S}}=\sum_{m=1}^{N}\epsilon _{m}|m\rangle \langle
m|+\sum_{n<m}^{N}V_{mn}(|m\rangle \langle n|+|n\rangle \langle m|),
\label{HamiltonianSystem}
\end{equation}%
where the Hilbert space basis states $|m\rangle $ denote the presence of
an excitation at the $m$th chromophore and $\epsilon _{m}$ are relative site
energies with respect to the chromophore with the lowest absorption energy.
The $V_{mn}$ can be due to Coulomb coupling of the transition densities (F%
\"{o}rster) or due to overlap of electronic wavefunctions (Dexter).
We denote the eigenbasis of the Hamiltonian (\ref%
{HamiltonianSystem}) as the exciton basis $|M\rangle
=\sum_{m}c_{m}(M)|m\rangle $, where $H_{\rm S}|M\rangle =\epsilon_{\rm M}|M\rangle $.
The multichromophoric system interacts with a thermal phonon bath. The
dominant component of the system-bath Hamiltonian is associated with site-energy
fluctuations \cite{Cho05,Adolphs06}, i.e.~
$H_{\mathrm{SB}}=\sum_{m}q_{m}|m\rangle \langle m|$,
where $q_{m}$ are operators describing the coupling
to the coordinates of the harmonic-oscillator bath.
The phonon terms $q_{m}|m\rangle \langle m|$ induce relaxation and
dephasing without changing the number of excitations. We assume that the
bath correlator can be simplified as $\langle q_{m}(t)q_{n}(0)\rangle
=C_{mn}\langle q(t)q(0)\rangle $.  $C_{mn}$ is a dimensionless time-independent factor
that takes into account the spatial correlations in the phonon bath.
For spatially uncorrelated environments it will simply be given by $C_{mn}=\delta_{mn}$.
In this work, we will also take into account a phenomenological model for
these correlations as will be explained later.
The time-dependent part of the correlator is the same for all sites \cite{Cho05}.
Additionally, there are two processes that lead to irreversible loss of the exciton \cite%
{Sener04,Leegwater96,Castro07,Mohseni08}. One is the excitation loss due to
recombination of the electron-hole pair. The other mechanism describes the excitation
transfer to the reaction center (acceptor) and subsequent trapping associated with the
charge separation event. These effects are taken into account by the anti-Hermitian Hamiltonians,
$-\mathrm{i}H_{\mathrm{recomb}}=-\mathrm{i}\hbar \Gamma \sum_{m}^{N}|m\rangle \langle m|$,
with $\Gamma $ the inverse lifetime of the exciton and $-\mathrm{i}H_{\mathrm{trap}}=-%
\mathrm{i}\hbar \sum_{m}^{N}\kappa _{m}|m\rangle \langle m|$, with
$\kappa _{m}$ the trapping rates at site $m$.

In summary, the dynamics of the reduced density matrix of the system
can be described by the Lindblad master equation in the Born-Markov
and secular approximations as \cite{BreuerBook}:
\begin{eqnarray}\label{MasterEquation}
\frac{d\rho (t)}{d t}&=&-\frac{\mathrm{i}}{\hbar }[H_{\mathrm{S}%
}+H_{\mathrm{LS}},\rho (t)]+ \mathcal{L}\rho (t) \\ \nonumber
 &&-\frac{1}{\hbar} \{ H_{\rm recomb}, \rho(t) \}
 -\frac{1}{\hbar} \{ H_{\rm trap}, \rho(t) \},
\end{eqnarray}
where $\{,\}$ denotes the anti-commutator. The right-hand side of
Eq.~(\ref{MasterEquation}) defines the superoperator $\mathcal{M}$.
$\mathcal{L}$ is the Lindblad superoperator derived from the phonon bath coupling,
\begin{eqnarray}
\mathcal{L}\rho(t) &=&\sum_{\omega,m,n }\gamma_{mn} (\omega )[A_{m}(\omega )\rho(t)
A_{n}^{\dagger }(\omega )  \label{LindbladSuperoperator} \\
&&-\frac{1}{2}A_{m}(\omega )A_{n}^{\dagger }(\omega )\rho(t) -\frac{1}{2}\rho(t)
A_{m}(\omega )A_{n}^{\dagger }(\omega )].  \nonumber
\end{eqnarray}
The sum runs over all possible transitions in the single exciton manifold
and all the sites. The Lindblad generators are
$A_{m}(\omega )=\sum_{\epsilon _{\rm M}-\epsilon_{\rm N}=\hbar \omega }c_{m}^{\ast }(M)c_{m}(N)|M\rangle \langle N|$,
where the
summation runs over all transitions with frequency $\omega $ in the
single-excitation manifold.
The Fourier transform of the bath correlation function leads to the rates
$\gamma_{mn} (\omega)=2\pi C_{mn} \lbrack J(\omega )(1+n(\omega ))+$ $J(-\omega
)n(-\omega )],$ where $n(\omega )$
is the bosonic distribution function at temperature $T$. Here, we assume an
Ohmic spectral density with\ $J(\omega )=0$ for $\omega <0$ and $J(\omega )=
\frac{E_{\mathrm{R}}}{\hbar \omega _{c}} \omega \exp (-\frac{\omega
}{\omega _{c}})$ elsewhere, with cutoff $\omega _{c}$, and reorganization
energy $E_{\mathrm{R}}=\hbar \int_{0}^{\infty }d\omega \frac{J(\omega )}{\omega }$
\cite{Cho05}. The pure dephasing part of Eq.~(\ref{LindbladSuperoperator}) is
obtained in the limit $\omega \to 0$. The Lindblad generators
are $A_{m}(0)=\sum_{M} |c_{m}(M)|^2 |M\rangle \langle M|$ and
the rate is $\gamma _{\phi,mn }=2\pi C_{mn} \frac{E_{\mathrm{R}}}{\hbar \omega _{c}}kT.$
The Lamb shift Hamiltonian \cite{BreuerBook}
is given by H$_{\mathrm{LS}}=E_{\rm R} \sum_{M,m} |c_{m}(M)|^4 |M\rangle \langle M|$,
where we only take into account the most significant, diagonal part,
see Ref.~\cite{Adolphs06} and supplementary material thereof.

The competition of trapping and recombination processes leads to the concept of the
energy transfer efficiency (ETE) as the integrated probability of
successful exciton entrapment by the reaction center
\cite{Sener04, Castro07, Mohseni08},
\begin{equation}
\eta =\frac{2}{\hbar }\int_{0}^{\infty }\mathrm{Tr}\{H_{\mathrm{trap}}\rho
(t)\}dt.  \label{Efficiency}
\end{equation}
The efficiency is suppressed by the lifetime of the excitation due to
exciton recombination. Quantitatively, the suppression is given by the
\textquotedblleft deficiency\textquotedblright, $\bar{\eta}=\frac{2}{\hbar }%
\int_{0}^{\infty }\mathrm{Tr}\{H_{\mathrm{recomb}}\rho (t)\}dt$. One finds the relation
$\eta +\bar{\eta}=1$ which implies that the excitation ultimately
either is trapped or recombines.

\section {Contributions to the energy transfer efficiency}

We partition the efficiency based on a Green's function method.
First, note that with Eq.~(\ref{LiouvilleQuantumWalk}) the efficiency can be simplified as
\begin{equation} \label{EfficiencyGreensfunction}
\eta =-\frac{2}{\hbar }\mathrm{Tr}\{H_{\mathrm{trap}}\mathcal{M}^{-1}\rho(0)\},
\end{equation}
The trapping and recombination Hamiltonians ensure that $\mathcal{M}$ is invertible
since non-trivial eigenstates with zero eigenvalue, such as the
usual thermal equilibrium, do not exist.
The Green's function interpretation of $-\mathcal{M}^{-1}$ can be readily confirmed by examining the
Laplace-transformed master equation.
Next, the superoperator is decomposed into $\mathcal{M=H}_{\mathrm{ref}}+%
\mathcal{R}$, where the recombination-trapping part $\mathcal{H}_{\mathrm{ref}}
\rho=-\frac{1}{\hbar }\{ H_{\rm recomb}, \rho(t) \} -\frac{1}{\hbar } \{ H_{\rm trap}, \rho(t) \}$,
 is taken as a
reference point and the remainder is given by $\mathcal{R}\rho=-\frac{
\mathrm{i}}{\hbar }[H_{\mathrm{S}}+H_{\mathrm{LS}},\rho ]+\mathcal{L}\rho$. This
allows us to express the Green's function exactly as \cite{MukamelBook},
\begin{equation}
\mathcal{M}^{-1}=\mathcal{H}_{\mathrm{ref}}^{-1}+\mathcal{H}_{\mathrm{ref}
}^{-1}\mathcal{R M}^{-1}.  \label{GreensfunctionSummation}
\end{equation}
The first term in (\ref{GreensfunctionSummation}) describes the irreversible
dynamics resulting from excitonic transfer to the acceptor and recombination of
the excitation. The second term accounts for the effect of
the Hamiltonian and the Lindblad operator. This term can
be decomposed into various physical processes such as coherent hopping or
phonon-mediated jumps from site-to-site, $\mathcal{R}=\sum_{k}\mathcal{R}_{k}$.
We use Eq.~(\ref{GreensfunctionSummation}) and the relation $\eta =1-\bar{\eta}$
to arrive at the partitioning of the efficiency, $\eta =\sum_{k}\eta _{k}$, with
\begin{equation}
\eta _{k}=-\frac{2}{\hbar }\mathrm{Tr}\{H_{\mathrm{recomb}}\mathcal{H%
}_{\mathrm{ref}}^{-1}\mathcal{R}_{k}\mathcal{M}^{-1}\rho(0)\}.
\label{GreensfunctionContribution}
\end{equation}
The part related to the first term in Eq.~(\ref{GreensfunctionSummation})
gives just one and thus cancels with the one from the conservation property.
Note that each $\eta _{k}$ is dependent on the initial state, the
overall dynamics $\mathcal{M}^{-1}$, and the particular process
$\mathcal{R}_{k}$ weighted by a diagonal trapping-recombination
operator
$H_{\mathrm{recomb}}\mathcal{H}_{\mathrm{ref}}^{-1}$. The
trace then takes into account the local effect of $\mathcal{R}_{k}$
on the population of all the site basis states $|m\rangle$.

\section{Application to the FMO complex}

The Fenna-Matthews-Olson
protein complex is a chromophoric trimer where each subunit consists
of seven chlorophylls embedded in a protein environment
\cite{Li97,Engel07,Mueh07}. It connects
the photosynthetic antenna with the reaction center in {\it Chlorobium tepidum}.
We use the seven-level Hamiltonian of the FMO complex as given in Ref.~\cite{Cho05}
and the bath spectrum described earlier with the reorganization energy $E_{%
\mathrm{R}}=35$cm$^{-1}$ and cutoff $\omega _{c}=150$cm$^{-1}$, inferred
from Fig.~2 of Ref.~\cite{Adolphs06}. The transfer of the excitation
from the FMO to the reaction center occurs via site 3 with the rate $\kappa
_{3}$, which is a free parameter in our simulations and, if not otherwise
stated, is taken to be $\kappa _{3}=1$ps$^{-1}$ \cite{Mohseni08}. The exciton lifetime is
assumed to be $1/\Gamma =1$ns \cite{Owens87}. 
The initial state for
the simulation is taken to be an unbiased classical mixture of all sites
except the trapping site 3.
At first, we will assume that the phonon modes at different chlorophylls
are not spatially correlated with each other, which is equivalent to setting
$C_{mn}=\delta_{mn}$. Later, we will relax this assumption and investigate
the contributions as a function of a phenomenological parameter, the
correlation length $R_c$.

\begin{figure*}
\includegraphics[scale=0.42]{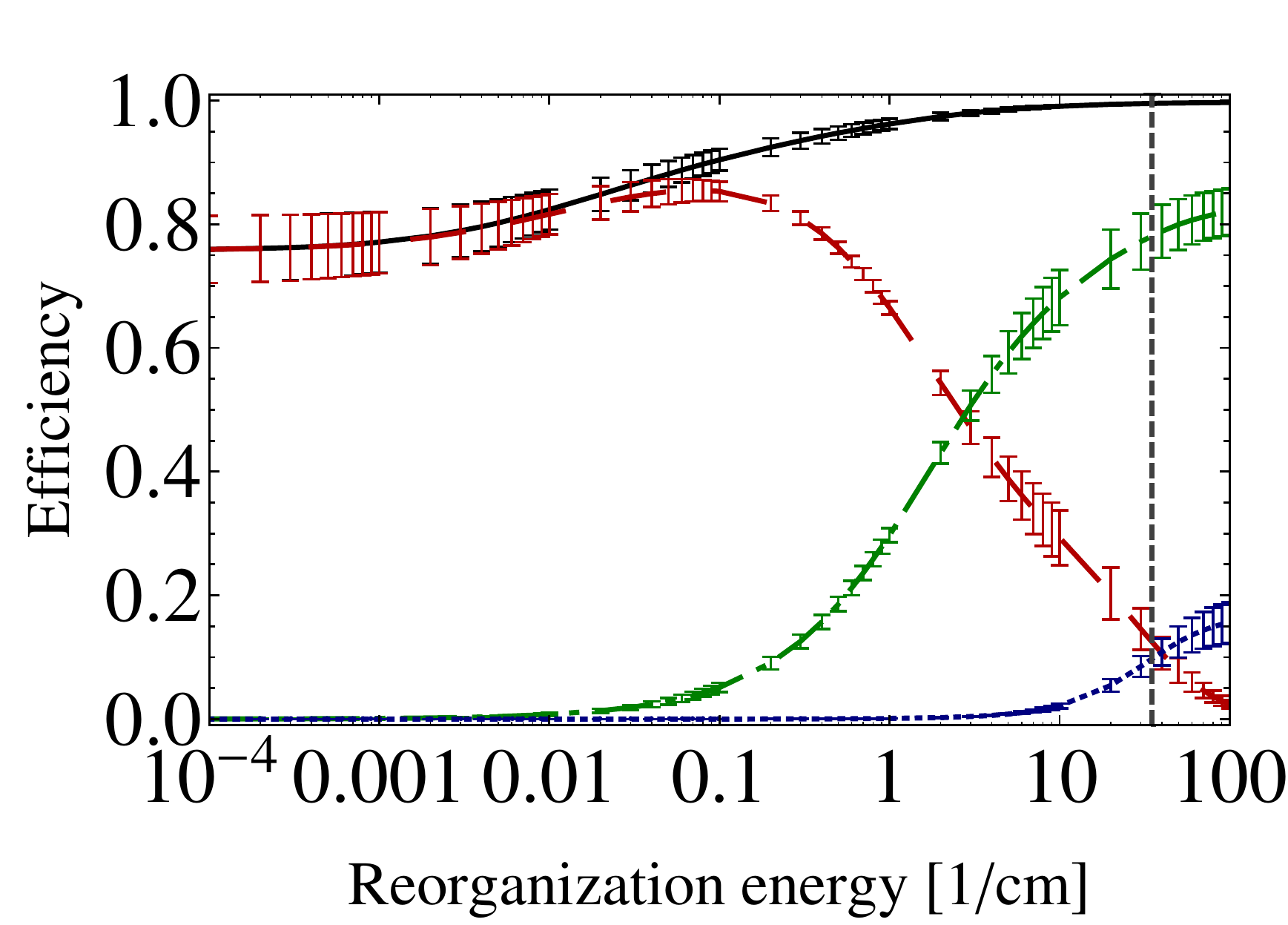} \hspace{1cm}
\includegraphics[scale=0.7]{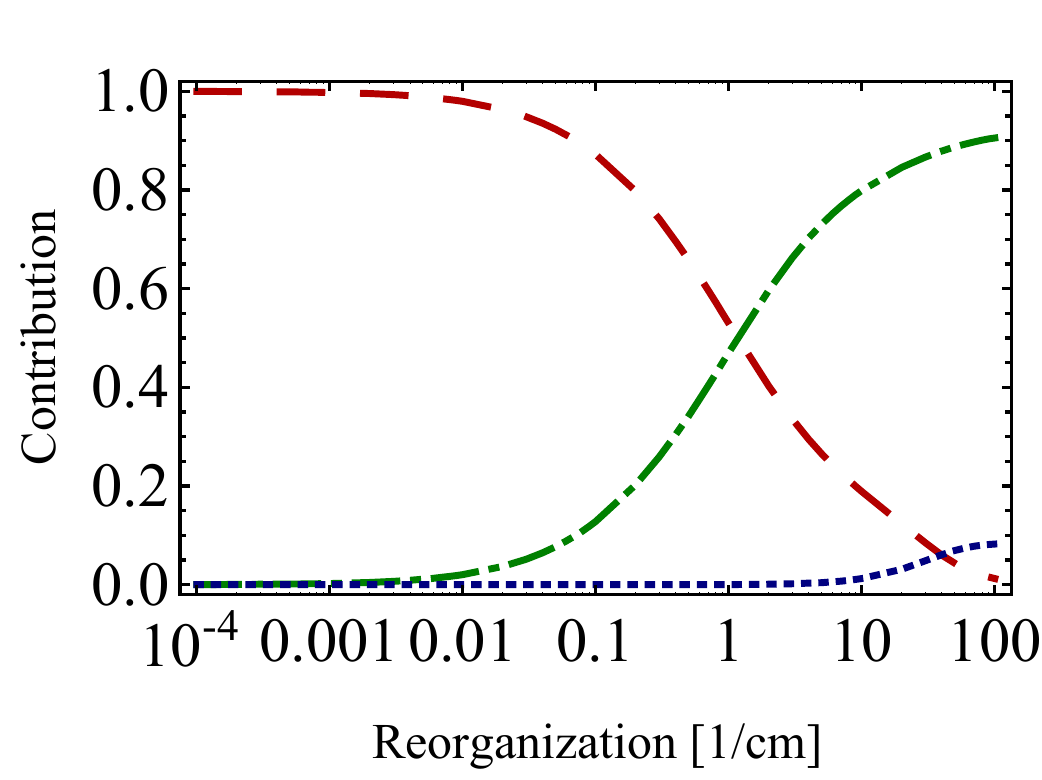}
\includegraphics[scale=0.42]{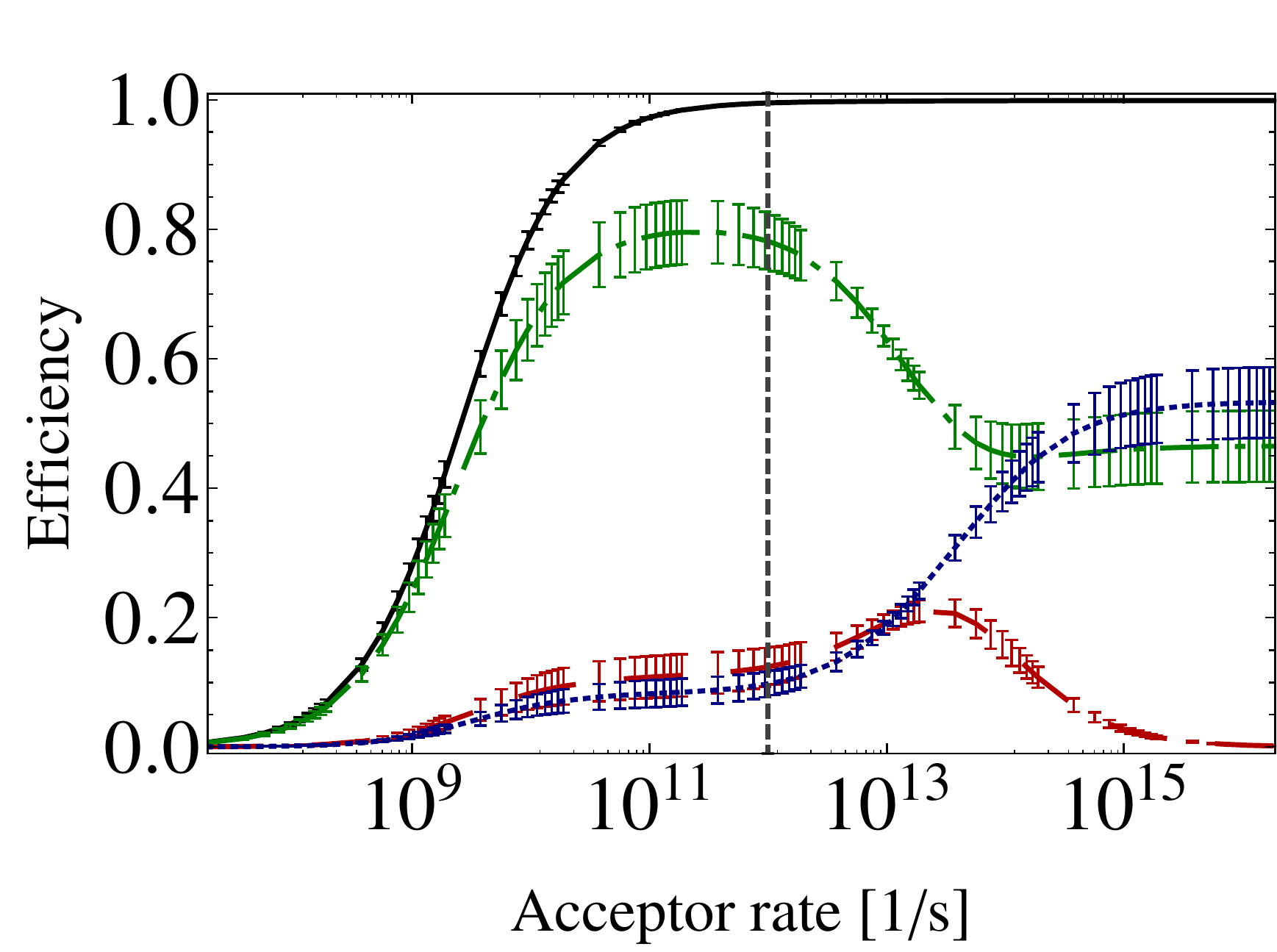}\hspace{1cm}
\includegraphics[scale=0.7]{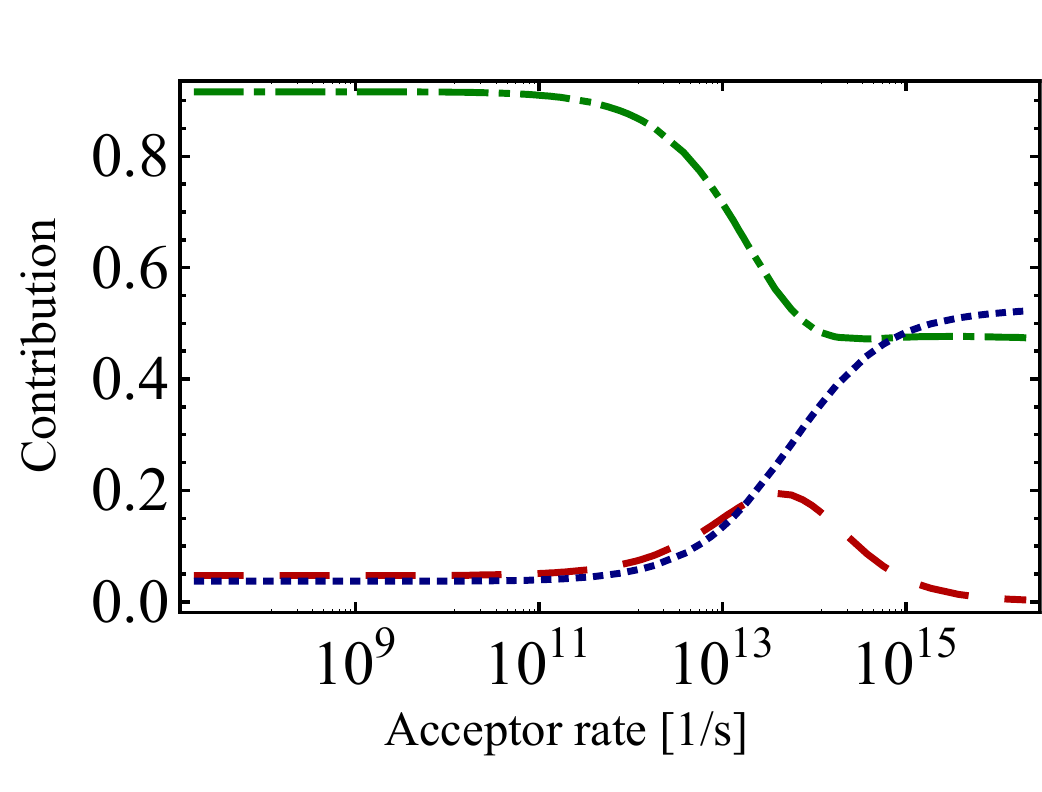}
\includegraphics[scale=0.42]{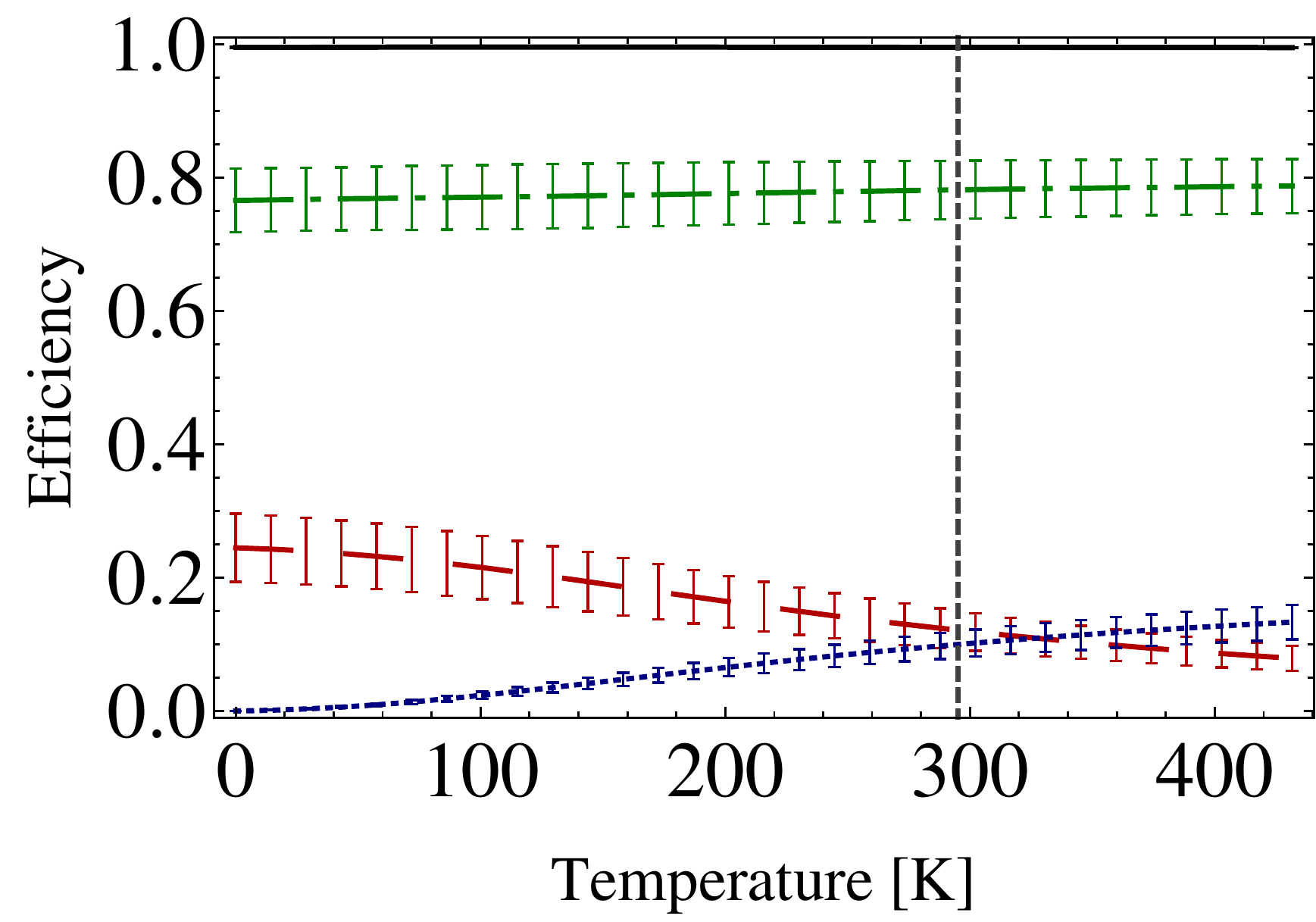}\hspace{1cm}
\includegraphics[scale=0.7]{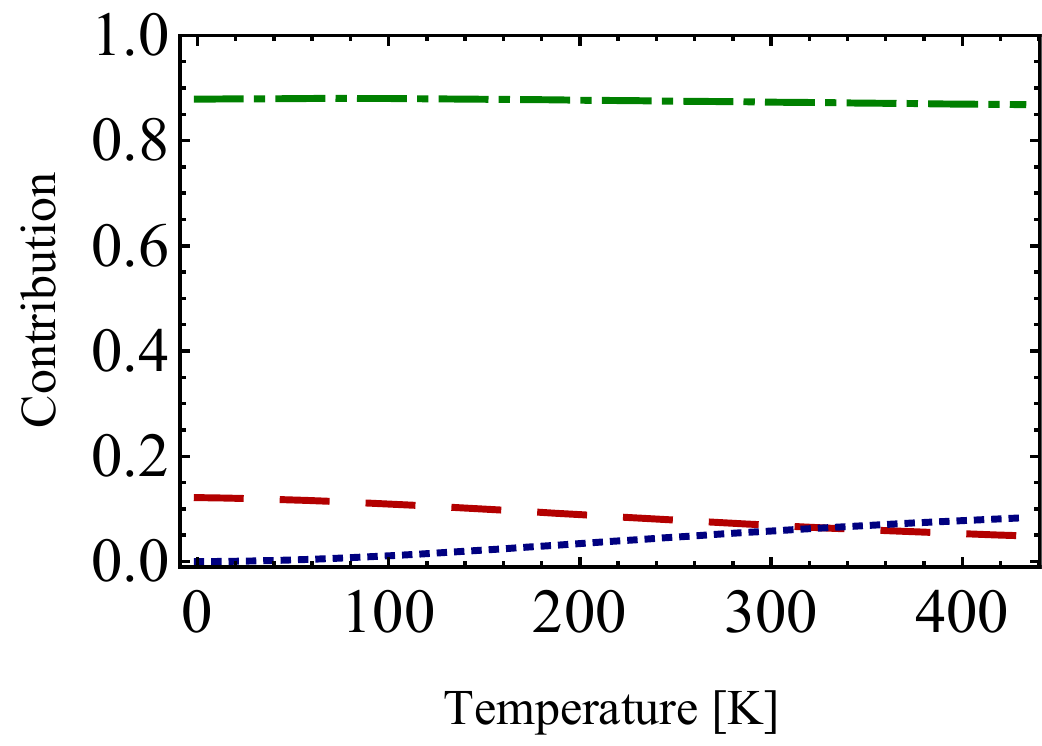}
\caption{Contributions of the basic physical processes to the overall
energy transfer efficiency of the Fenna-Mathews-Olson protein complex according to
Eq.~(\protect\ref{GreensfunctionContribution}) (left panels) and Eq.~(\ref{ContributionAveraged})
with normalization (\ref{ContributionAveragedNormalization}) (right panels).
The physical processes are the Hamiltonian evolution (red, $--$),
exciton relaxation (green, $-\cdot $), and exciton dephasing (blue, $\cdot \cdot \cdot $).
They add up to the efficiency (black, $-$) (left) or $1$ (right).
Depicted are the contributions as a function of a) reorganization
energy, essentially the overall strength of the phonon-bath coupling
b) the transfer rate to the reaction center, and c) the
temperature.
We assume that the initial state is an unbiased
classical mixture of all the sites of the FMO complex except target
site 3.
Typical parameters (shown as vertical lines) are $%
E_{\mathrm{R}}=35$cm$^{-1}$, $T=295$K, $\protect\kappa
_{3}=1$ps$^{-1}$, and $\Gamma =1$ns$^{-1}$ \protect\cite{Adolphs06}.
Using these parameters, the first (second) measure on the left (right) gives an estimated contribution
of 80\% (87.5\%) for exciton relaxation and 10\% (7.5\%) for the Hamiltonian evolution
at room temperature.
} \label{figContributionsBasic}
\end{figure*}

In Fig.~\ref{figContributionsBasic} (left panels) we present the contributions to
the ETE as a function of three relevant system parameters for the
three fundamental physical processes: Hamiltonian evolution with $H_{\rm S}$
and exciton relaxation and dephasing, both contained in $\mathcal{L}$.
The parameters explored are the
reorganization energy $E_{\mathrm{R}}$, which is a linear prefactor of all
the decoherence rates, the transfer rate to the reaction center
$\kappa _{3}$, and temperature $T$. In Ref.~\cite{Mohseni08}, we
reported the overall behavior of the efficiency with respect to
these parameters, especially the enhancement of the ETE by about 20\%
compared to no phonon-bath, $E_{\rm R}=0$. Even more significantly, the transport time of
the excitation to the reaction center reduces from over $50$ps to under $5$ps.
Here, by providing a more
detailed analysis of the underlying processes, we observe
a crossover from a purely quantum regime to a relaxation-dominated regime
with increasing phonon bath coupling strength $E_{\rm R}$,
see Fig.~\ref{figContributionsBasic} (a).
At room temperature and $E_{\mathrm{R}}=35$cm$^{-1}$
the purely quantum mechanical contribution of coherent hopping due
to the Hamiltonian is around 10\%. This is the main result of this work
and is to be understood within the context of our model and its approximations.
The major contribution is due to exciton relaxation induced by
the phonon-bath coupling. Site 3 has a large participation in the lowest-energy exciton:
relaxation helps the transport from the energetically higher initial state towards that site.
Dephasing processes have a contribution of around 8\% (see explanation below).
To summarize, the results so far indicate that the process of
energy relaxation determines the high efficiency and the fast transfer times
f the Fenna-Matthews-Olson protein complex and thus is essential to the
biological function of this system. Supporting this statement is a recent study that finds that
the protein environment induces a static red-shift of the energy of site 3 \cite{Mueh07}.

Slow movement of the protein scaffold will lead to static
disorder in the Hamiltonian and thus to inhomogeneous
broadening of spectral lines. We obtain the effect of static disorder
on the efficiency and the contributions by averaging over diagonal disorder
of the site energies $\epsilon_m$.
We assume that the site energies are normal distributed around the
Hamiltonian given in \cite{Cho05}
and with a site-dependent FWHM $\sigma_m$. Based on the
supplementary material of Ref.~\cite{Mueh07}
we choose $\sigma_{1,3,4}=60 $cm$^{-1}$, $\sigma_{2}=100 $cm$^{-1}$,
and $\sigma_{5,6,7}=120 $cm$^{-1}$. These values lead to good agreement
of measured and simulated spectra and can be explained by
the different amounts of mobile water molecules in the vicinity of the
different chlorophylls of the FMO protein complex.
In Fig.~\ref {figContributionsBasic} the resulting distributions of
the efficiency and the contributions are depicted as error bars.
At small $E_{\rm R}$ the efficiency and the prevalent quantum mechanical contribution
have a broadening of around 10\%. This is due to the fact
that the static disorder changes the energy difference of interacting sites which
can facilitate or inhibit transport.
At larger $E_{\rm R}$ the efficiency has a small broadening of less than 1\%.
In the presence of an environment the system is more robust against static disorder.
A smaller contribution of the quantum mechanical part is compensated by a larger
contribution of relaxation and vice versa.

Fig.~\ref {figContributionsBasic} (b) shows the dependence of the contributions
on the transfer rate to the acceptor. The behavior
can be explained by the characteristic time scales of the various processes.
At room temperature and $E_{\mathrm{R}}=35$cm$^{-1}$, the relaxation
processes occur on a time scale of 1ps and are most important when
the acceptor rate is of the same order of magnitude. The
Hamiltonian and dephasing act on faster time scales such that
their contributions are stronger at higher acceptor rate values.
The contribution of dephasing can be understood from
the fact that, within the Redfield/Lindblad treatment, dephasing acts in the energy basis:
an initial state localized at a site dephases to a state that has overlap with the target site.
For example, the initial state $\rho(0)=\vert 1\rangle \langle 1 \vert=\sum_{\rm M,N}
c_{1}(M) c_{1}^{\ast}(N) \vert M\rangle \langle N\vert$
evolves in the presence of only dephasing to
$\rho(t \to \infty) = \sum_{\rm M}  |c_{1}(M)|^{2} \vert M\rangle \langle M\vert$,
which has $\langle 3| \rho(t \to \infty) |3\rangle \neq 0$.

Finally, Fig.~\ref {figContributionsBasic} (c) shows the temperature
dependence of the ETE contributions at $E_{\rm R}=35$cm$^{-1}$.
The role of the Hamiltonian is about 20\% at zero temperature.
The relaxation processes are only weakly dependent on $T$:
the main contribution is the temperature-independent spontaneous
emission of energy into the phonon bath, leading to energy funneling
towards the site with the lowest energy, site 3.
The behavior of the dephasing processes is explained by $\gamma _{\phi }\sim kT$.
In all figures, the contribution of the Lamb
shift Hamiltonian is less than 1.5\% and thus not depicted.

\begin{figure}
\includegraphics[scale=0.42]{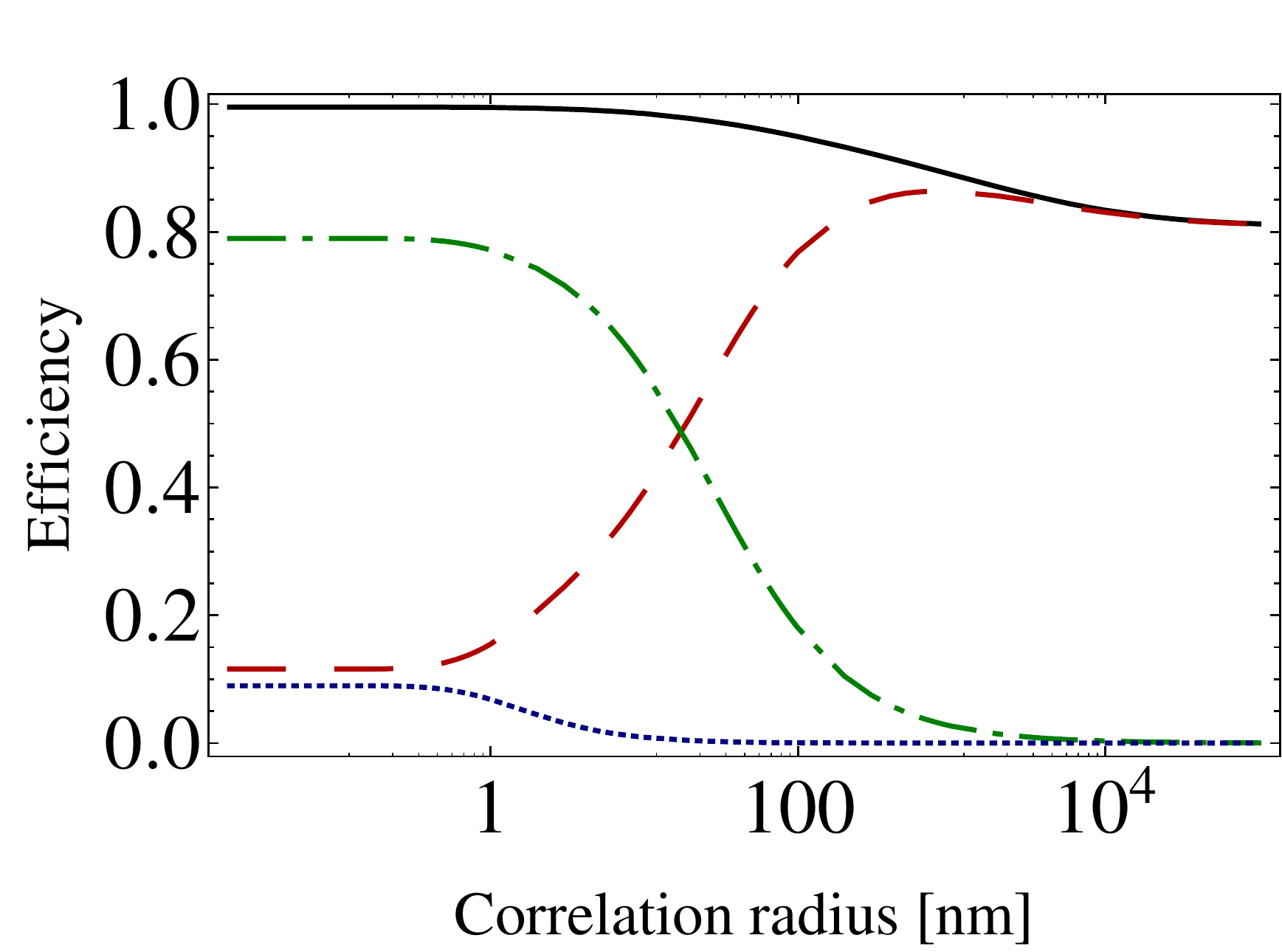}
\includegraphics[scale=0.42]{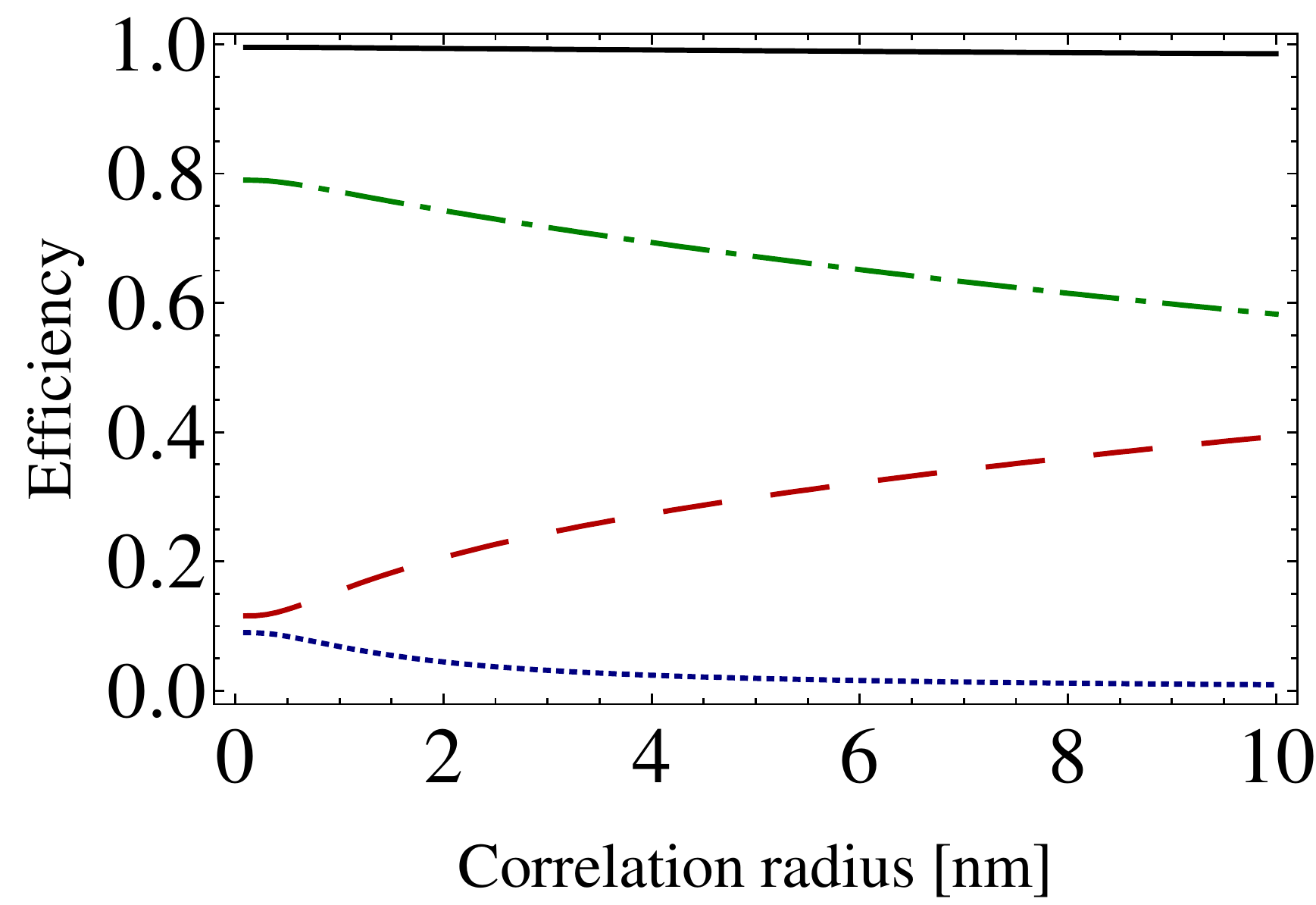}
\caption{
Efficiency and contributions of the basis physical processes as a function
of the correlation radius in the environment in the Fenna-Matthews-Olson complex.
The parameters used are $E_{\mathrm{R}}=35$cm$^{-1}$, $T=295$K,
$\kappa_{3}=1$ps$^{-1}$, and $\Gamma =1$ns$^{-1}$.
The lower panel is a magnification of the physically relevant region of the upper panel.
} \label{figContributionsCorrelation}
\end{figure}

Up to this point we assumed that the site energy fluctuations of a particular site
caused by the coupling to a vibrational bath are uncorrelated from
the fluctuations at another site. We relax this assumption to
take into account spatially correlated fluctuations present in
realistic chromophoric systems embedded in a protein environment.
Specifically, instead of an uncorrelated environment with $C_{mn}=\delta_{mn}$ we
include correlations with $C_{mn}=e^{-R_{mn}/R_c}$ \cite{Adolphs06}.
$R_{mn}$ is the intermolecular distance between chlorophyll
$m$ and $n$ and $R_c$ is the correlation radius. In the limit of
$R_c \to 0$ one has the uncorrelated case. In the limit of
$R_c \to \infty$ all fluctuations are perfectly correlated and thus lead only to
an irrelevant global phase in the exciton dynamics: this limit is equivalent to no phonon
bath at all. In Fig.~\ref{figContributionsCorrelation} we plot the efficiency
and the contributions as a function of the correlation radius
for the Fenna-Matthews-Olson protein complex.
We recover the two limits: at small correlation radius the efficiency and
the contributions are the same as in Fig.~\ref{figContributionsBasic}.
For large correlation radius the quantum coherence contribution is dominant
and there is no contribution of relaxation and dephasing.
In between, we observe a crossover which happens when the correlation
radius is of the order of the intermolecular distance.

\section{Beyond local contributions}

The method for partitioning the
efficiency into contributions presented in the last section sheds light on
the role of the basic physical processes in the overall open systems
dynamics.
Our approach resembles the sojourn expansion used in Ref.~\cite{Sener04}
for a classical random walk and the method used in Ref.~\cite{Leegwater96}
for studying incoherent versus coherent energy transfer in the
Haken-Strobl model.
However, it shows a certain local feature:
only processes that are directly connected to site population
elements play a role. This can be seen in Eq.~(\ref%
{GreensfunctionContribution}) from the fact that $\mathcal{H}_{\mathrm{ref}%
}^{-1}$ and $H_{\mathrm{recomb}}$ are diagonal operators.
Thus the contributions of all $\mathcal{R}_{k}$ which do not lead to
transfer to a site population vanish in the trace, e.g., in the case of coherence to
coherence processes. Note that this effect would have been pronounced
had we not used the identity $\eta=1-\bar{\eta}$ before
Eq.~(\ref{GreensfunctionContribution}). In this case only
those $\mathcal{R}_{k}$ that create overlap with the target site 3 would
contribute. To overcome these issues, we develop our complementary measure motivated by the concept of
energy transfer susceptibilities, discussed in \cite{Mohseni08}.

In this method, we partition the
efficiency into terms involving the variational change of the
excitation trapping probability density $\frac{2}{\hbar}
\mathrm{Tr}\{H_{\mathrm{trap}}\rho (t)\}$ with respect to the
different physical processes. We first provide a representation for
a quantum master equation in the Markov approximation that is
tailored to the problem at hand:
\begin{equation}
\frac{d}{d t}\rho(t)=\sum_{k}\frac{1}{t}\int_{0}^{t}\mathcal{F}
(t,t^{\prime })\mathcal{M}_{k}\mathcal{F}(t^{\prime },0)\rho
(0)dt^{\prime }.  \label{TrajectoryMasterEquation}
\end{equation}
We used the decomposition $\mathcal{M}=\sum_{k}\mathcal{M}_{k}$
and the quantum map $\rho(t)=\mathcal{F}(t,0)\rho(0)$.
This equation is equivalent to Eq.~(\ref{LiouvilleQuantumWalk}),
but the individual terms in the sum are different from just $\mathcal{M}_{k}$:
the integrand can be understood as the realization of a non-unitary quantum walk in
Liouville space. The quantum walk starts at the initial state $\rho(0)$%
, evolves with $\mathcal{F}$ for a time interval of $[0,t^{\prime }]$, then
a perturbation $\mathcal{M}_{k}$ is applied at an arbitrary time $t^{\prime }
$, after which it undergoes a post-evolution until time $t$. The second
evolution $\mathcal{F}(t,t^{\prime })$ leads to quantum interference
between $\mathcal{M}_{k}$ and other generators. Integrating over
all possible paths gives the average effect of the $k$th
generator to the time variation of the density matrix.

Eq.~(\ref{TrajectoryMasterEquation}) can be reexpressed by introducing scalar
dimensionless quantities $\lambda _{k}$ associated with each term $\mathcal{M}_{k}$,
such that $\mathcal{M}_{k}\rightarrow $ $\lambda _{k}\mathcal{M}_{k}$ in the
neighborhood $\lambda_{k}\rightarrow $ $1$. This leads to the
master equation $\frac{d}{d t} \rho(t) = \frac{1}{t} \sum_k
\frac{\partial}{\partial \lambda_k} \rho(t)$, which is equivalent to Eq.~(\ref{TrajectoryMasterEquation}).
We employ it to partition the ETE in Eq.~(\ref{Efficiency}) into contributions given by
\begin{equation}
\eta _{k}= \frac{2}{\hbar} \int_{0}^{\infty }dt\int_{0}^{t}dt^{\prime }\frac{1}{%
t^{\prime }}\frac{\partial }{\partial \lambda _{k}}\mathrm{Tr}\{H_{%
\mathrm{trap}}\rho (t^{\prime })\}.  \label{ContributionAveraged}
\end{equation}
We assume that the initial state $\rho (0)$ has no overlap with the trapping
sites. The terms $\frac{2}{\hbar} \frac{\partial }{\partial \lambda _{k}}\mathrm{Tr}%
\{H_{\mathrm{trap}}\rho (t)\}$ can be interpreted as the susceptibility with
respect to the process $\mathcal{M}_{k}$ of the exciton trapping probability
density $\frac{2}{\hbar} \mathrm{Tr}\{H_{\mathrm{trap}}\rho (t)\}$ at time $t$. The double
integration is then to be considered as a time averaging of this quantity.
The factor $\frac{1}{t}$ in the integrand arises from Eq.~(\ref{TrajectoryMasterEquation}).
We solve Eq.~(\ref{ContributionAveraged}) by numerical integration.
The resulting contributions $\eta_{k}$ are not normalized, meaning
that $0\le \eta_k \le 1$ does not necessarily hold.
To define a proper normalization we separate the $\eta_k$ into positive and negative terms,
i.e.~$\eta =\eta^{+} + \eta^{-} = \sum_{k_{+}}\eta _{k_{+}}+\sum_{k_{-}}\eta_{k-}$
where one has $\eta _{k_{+}}>0$ and $\eta_{k_{-}}<0$
and the overall positive or negative contributions $\eta ^{\pm}$. 
Now one can introduce a normalization given by:
\begin{equation}
\widetilde{\eta }_{k_{\pm }}=\frac{\eta _{k_{\pm }}}{\eta ^{\pm }}.
\label{ContributionAveragedNormalization}
\end{equation}
This allows us to interpret $\widetilde{\eta }_{k_{\pm }}$ as positive or negative
percentage contributions to the overall positive or negative contribution.
The normalized contributions $\widetilde{\eta }_{k_{\pm }}$
show similar behavior for the basic processes as the Green's function measure
Eq.~(\ref{GreensfunctionContribution}), see Fig.~\ref{figContributionsBasic} (right panels).
This can be seen as evidence that, one the one hand, the normalization procedure was appropriate
and, on the other hand, that despite its local feature the Green's function measure
obtains consistent results.
The second measure generally assigns less contribution to the
free Hamiltonian, about 7.5\% at $E_{\rm R}=35$cm$^{-1}$ and $T=295K$.
For these parameters, the contribution of relaxation is about 87.5\%
and dephasing about 5\%. At zero temperature the Hamiltonian contribution
is 12.5\%.

\begin{figure}[tbp]
\includegraphics[scale=1.00]{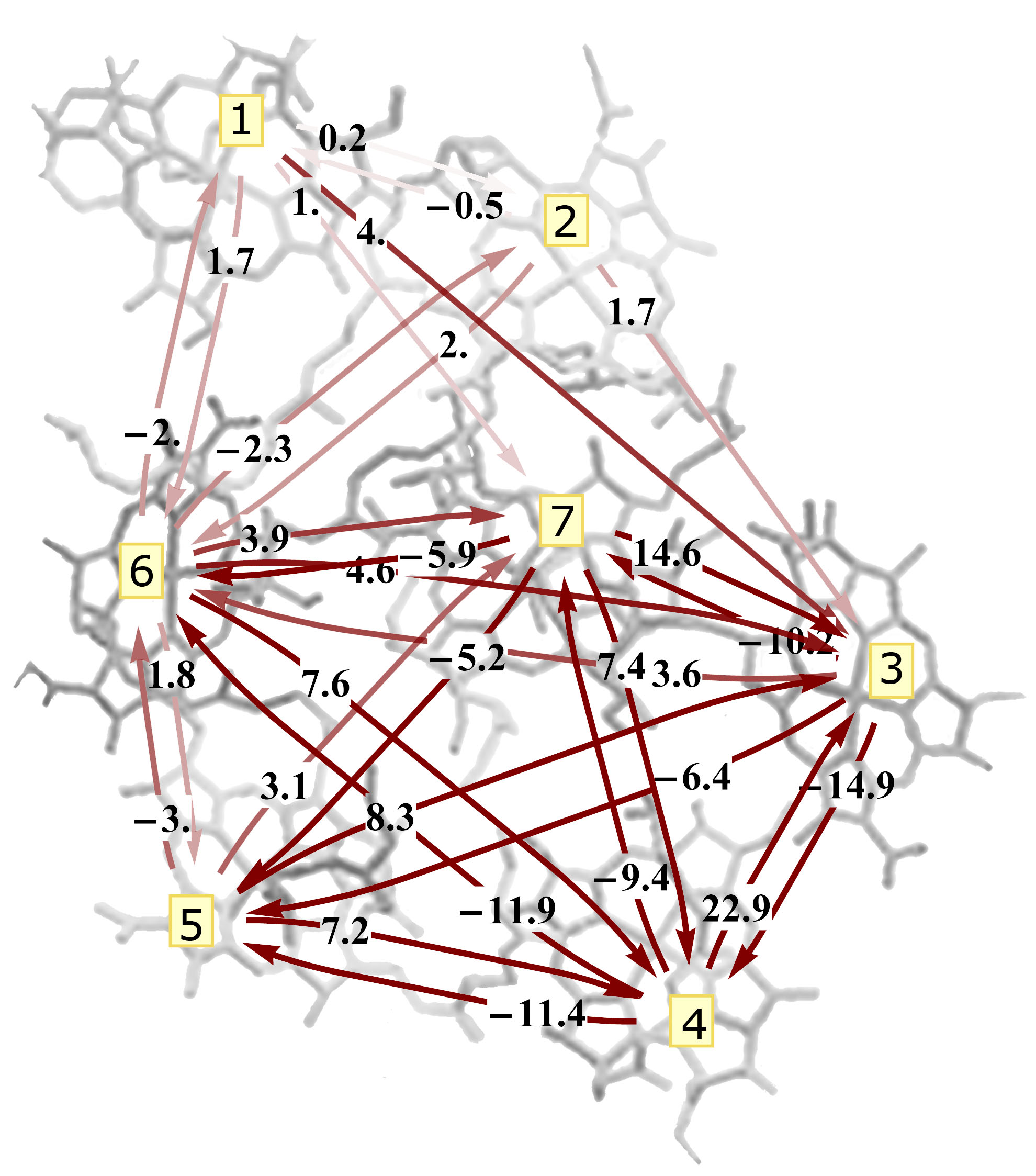}
\caption{The Fenna-Matthews-Olson protein complex and
the contributions in percent of the relaxation pathways to the overall
relaxation contribution of the energy
transfer efficiency, using Eq.~\protect\ref{ContributionAveraged} and the normalization
Eq.~\protect\ref{ContributionAveragedNormalization}.
For clarity some contributions below 2\% are not depicted.
The initial state is a classical mixture of all sites except target site 3.
Standard parameters are $E_{\mathrm{R}}=35$cm$^{-1}$, $T=295$K, $\protect\kappa _{3}=1
$ps$^{-1}$, and $\Gamma =1$ns$^{-1}$. The contributions show directionality
in their sign and reveal the important pathways. }
\label{figContributionsSiteToSite}
\end{figure}

The measure based on susceptibilities can be used to quantify the contributions of exciton
relaxation pathways in the site basis. Formally, we look at the contribution
of all site-to-site jumps and the corresponding damping of diagonal
populations, see Ref.~\cite{Mohseni08}. In Fig.~\ref%
{figContributionsSiteToSite} we show the contributions of the various
pathways for the Fenna-Matthews-Olson protein complex when the system is
initially in a classical mixture of all sites except the target site 3.
Jumps toward the target site 3 contribute positively while jumps away from
the target site contribute negatively. Large contributions come from nearest
neighbor jumps, and site 4 and 7 are revealed as hubs in the transfer
toward site 3\textbf{.}

\section{Conclusion}

In this work, we have addressed the role of
quantum coherence and the environment in excitonic energy transfer.
To this end, we have characterized the underlying processes
constituting the open quantum walk of the excitation in terms of
their contribution to the transfer efficiency. The methods presented
here are general and can be applied to a large class of transport systems
in the presence of Markovian environments. Within both the Green's function and the
energy transfer susceptibility formalisms we conclude that the major
part of the high efficiency of the Fenna-Matthews-Olson protein
complex of about 80\% or 87.5\% is due to environment-induced relaxation
down to the lowest energy site. The role of quantum
coherence induced by the Hamiltonian dynamics can be quantified
at around 10\% or 7.5\% respectively. Furthermore, we used the
susceptibility measure to assign percentage-wise contributions to
exciton relaxation pathways in the molecular basis.
The detailed analysis of the open quantum dynamics presented in this work
could be harnessed for engineering artificial materials such as quantum
dots \cite{Nazir05} to achieve optimal energy transport in realistic environments.

\begin{acknowledgments}
We would like to acknowledge useful discussions with
G.R.~Fleming, S.~Lloyd, and A.T.~Rezakhani. We
thank the Faculty of Arts and Sciences of Harvard University, the
Army Research Office (project W911NF-07-1-0304), and Harvard's
Initiative for Quantum Science and Engineering for funding.
\end{acknowledgments}

\end{document}